\documentclass[aps,twocolumn,english,superscriptaddress,citeautoscript,preprintnumbers,amsmath,amssymb,floatfix,footinbib,prb]{revtex4}
\usepackage{hyperref}
\hypersetup{colorlinks=true,linkcolor=red,citecolor=blue}
\usepackage{amsmath}
\usepackage{graphicx}
\usepackage{dcolumn}
\usepackage{float}

\begin{document}
\title{Multigap Superconductivity in a charge density wave superconductor LaPt$_2$Si$_2$}
\author{Debarchan Das}
\thanks{Present address: Institute of Low Temperature and Structure Research, Polish Academy of Sciences, Wroc{\l}aw 50-950, Poland}
\affiliation{Department of Physics, Indian Institute of Technology, Kanpur 208016, India}
\author{Ritu Gupta}
\affiliation{Department of Physics, Indian Institute of Technology, Kanpur 208016, India}
\author{A. Bhattacharyya}
\email{amitava.bhattacharyya@rkmvu.ac.in}
\address{Department of Physics, Ramakrishna Mission Vivekananda Educational and Research Institute, Belur Math, Howrah 711202, West Bengal, India}
\author{P. K. Biswas}
\affiliation{ISIS Facility, STFC, Rutherford Appleton Laboratory,Chilton, Oxfordshire, OX11 0QX, United Kingdom}
\author{D. T. Adroja}
\affiliation{ISIS Facility, STFC, Rutherford Appleton Laboratory,Chilton, Oxfordshire, OX11 0QX, United Kingdom}
\affiliation{Highly Correlated Matter Research Group, Physics Department, University of Johannesburg, PO Box 524, Auckland Park 2006, South Africa}
\author{Z. Hossain}
\email{zakir@iitk.ac.in}
\affiliation{Department of Physics, Indian Institute of Technology, Kanpur 208016, India}

\date{\today}

\begin{abstract}

The superconducting gap structure of a charge density wave (CDW) superconductor LaPt$_2$Si$_2$ ($T_c$ = 1.6~K) having a quasi two dimensional crystal structure has been investigated using muon spin rotation/relaxation ($\mu$SR) measurements carried out in transverse field (TF), zero field (ZF) and longitudinal field (LF) geometries. Rigorous analysis of TF-$\mu$SR spectra in the superconducting state corroborates that the temperature dependence of the effective penetration depth, $\lambda_L$, derived from muon spin depolarization, fits to an isotropic $s+s-$wave model suggesting that the Fermi surface contains two gaps of different magnitude rather than an isotropic gap expected for a conventional $s-$wave superconductor. On the other hand, ZF $\mu$SR data do not show any significant change in muon spin relaxation rate above and below the superconducting transition indicating the fact that time-reversal symmetry is preserved in the system. 

\end{abstract}

\maketitle

BCS theory\cite{Cooper,Bardeen1, Bardeen2} which explains superconductivity in the conventional systems, fails to unfold the mystery of witnessing superconductivity in some materials which form a new class of superconductors (SC), collectively classified as unconventional SC. This encompasses a variety of materials which includes cuprate, heavy-fermion superconductor, pnictide etc~\cite{Bednorz, Steglich, White, Kamihara, Johnston}. Unlike conventional SC where pairing is mediated by lattice vibrations or phonos, in unconventional SC, fluctuations of the order parameters play a crucial role in the formation of a superconducting ground state. Hence, the search for unconventional SC and understanding their pairing mechanism have become an intensely studied active research area for the past few decades. In this quest, discovery of superconductivity by suppressing spin density wave (SDW) ordering in Fe-based pnictides has received a considerable attention\cite{Kamihara, Johnston, Takahashi, Jeevan}. Both spin fluctuations and density fluctuations (associated with structural transition) are believed to be important in governing superconductivity in the system. Very recently, charge density wave (CDW) systems which can be recognized as nonmagnetic analogue of Fe-based pnictides, has gained significant research interest as the fluctuation associated with CDW is believed to be a key factor in inducing superconductivity in the system\cite{Imre, Kudo, Nagano,Gruner}.

\par

A new class of intermetallic series, RPt$_2$Si$_2$ (R=La, Pr) has been intensively investigated recently which exhibits strong interplay between CDW and superconductivity\cite{Nagano,Ritu1, Ritu2,Anand,Kumar}. RPt$_2$Si$_2$ crystalizes in primitive tetragonal CaBe$_2$Ge$_2$ type structure (space group $ P4$/$nmm$) \cite{Nagano,Ritu1, Ritu2} having a close resemblance to the ThCr$_2$Si$_2$ type structure found in pnictide and heavy fermion SC. However, one striking difference between these two structures accounts the fact that the former one lacks inversion symmetry in the crystal structure which contains two inequivalent [Ge1-Be2-Ge1] and [Ge2-Be1-Ge2] layers with Ca atom (or R atom for RPt$_2$Si$_2$) being sandwiched between them\cite{Eisenmann}. In this context it is to be mentioned that other than RPt$_2$Si$_2$, SrPt$_2$As$_2$ is another example which crystalizes in this crystal structure and exhibits coexistence of superconductivity and CDW \cite{Imre, Kudo, Kawasaki}. Moreover, this special feature in crystal structure is reminiscent of non-centrosymmetric SC\cite{Kimura, Pecharsky, Hillier, Bauer, Smidman} where the lack of inversion symmetry results in non uniform lattice potential which in turn creates an asymmetric spin orbit coupling allowing mix pairing symmetry between a spin singlet and  a spin triplet cooper pairs. Mixing of spin singlet and triplet pairing makes these non-centrosymmetric SC more likely to exhibit time reversal symmetry (TRS) breaking and the physics of the system can be modified by this broken symmetry. TRS breaking is rare and has only been observed directly in a few unconventional superconductors, e.g., Sr$_2$RuO$_4$~\cite{Luke}, UPt$_3$~\cite{gml}, (U;Th)Be$_{13}$~\cite{rhh}, (Pr;La)(Os;Ru)$_4$Sb$_{12}$~\cite{ya}, PrPt$_4$Ge$_{12}$~\cite{am}, LaNiC$_2$~\cite{ad1}, LaNiGa$_2$~\cite{ad2},  Re$_6$Zr~\cite{rps} and (Lu,Y)$_5$Rh$_6$Sn$_{18}$\cite{Bhattacharyya1,Bhattacharyya2}. Zero field muon spin relaxation (ZF$-\mu$SR) is a powerful tool to search for very weak TRS breaking fields or spontaneous internal field below $T_{\bf c}$. The presence of such low internal field limits the pairing symmetry and mechanism responsible for unconventional superconductivity. However, it is well established that this mixing of spin states does not always indicate TRS breaking\cite{Quintanilla, Bauer2}.

\par

In this framework, LaPt$_2$Si$_2$ turns out to be quite an interesting system hosting diverse exciting phenomena such as lack of inversion symmetry in crystal structure, CDW transition, structural phase transition from tetragonal to orthorhombic structure and superconductivity\cite{Nagano, Ritu1, Ritu2,Kubo}. Through small angle electron diffration study \cite{Nagano}, CDW wave vector has been confirmed to be ($n/3$,~$0,~0$), where $n$ (= 1, 2) is the order of reflection, which requires tripling of unit cell below T$_{CDW}$. Furthermore, theoretical prediction of coexistence of CDW and superconductivity in LaPt$_2$Si$_2$ by Kim $et~al.$ \cite{Kim} has been confirmed experimentally in our earlier reports\cite{Ritu1, Ritu2}. On the other hand, electronic structure calculation predicts quasi-two-dimensional nature of the Fermi surface\cite{Hase} similar to that seen for iron pnictides. These observations conjointly hint towards an exotic origin of superconductivity in the system. However, the superconducting gap structure, which is intimately related to the superconducting mechanism, remains unexplored till now. It requires microscopic techniques in order to have a proper understanding of the superconducting phase which emerges in presence of a competing CDW phase. This  motivates us to perform $\mu$SR experiments in the superconducting state to unveil the gap structure in LaPt$_2$Si$_2$. $\mu$SR is a dynamic method to resolve the type of pairing symmetry in superconductors\cite{Sonier}. In case of a type-II SC, the mixed or vortex state gives rise to a spatial distribution of local magnetic fields influencing the $\mu$SR signal through a relaxation of the muon spin polarization. In this Rapid Communications, we present the results of our detailed $\mu$SR investigation performed on LaPt$_2$Si$_2$ compound. Our results manifest the existence of multigap superconductivity in this system.

\subparagraph*{}

High quality  polycrystalline sample of LaPt$_2$Si$_2$ was prepared by arc melting the constituent elements taken in stoichiometric amount on a water cooled copper hearth in argon atmosphere, followed by annealing at 1000~$^{\circ}$C for a week. The detail procedure of sample preparation can be found in Ref \cite{Ritu1}. The detail of sample  characterization has been provided in Supplemental Material\cite{Das}. The $\mu$SR experiments were performed in MUSR spectrometer at ISIS pulsed muon facility of the Rutherford Appleton Laboratory, United Kingdom \cite{Lee}. The $\mu$SR measurements had been carried out in transverse-field (TF), zero-field (ZF) and longitudinal-field (LF) configurations. The powdered sample was mounted on a high purity silver plate using diluted GE varnish and covered with a thin Ag foil which was cooled down to 50~mK in a commercial dilution refrigerator (ICE). 100\% spin-polarized muon pulses were implanted into the sample and positrons from the resulting decay were collected in the detectors. TF- $\mu$SR experiments were carried out in the superconducting mixed state under different applied fields ranging from 100~G to 300~G. TF-$\mu$SR measurements were performed in the field cooled mode in which the magnetic fields were applied above the superconducting transition temperature and the sample was then cooled down to 50 mK. For ZF-$\mu$SR measurements, the sample was cooled down to base temperature in true zero field. Data were collected while warming the sample. $\mu$SR data were analyzed using the free software package WiMDA\cite{Pratt}.

\subparagraph*{}

\begin{figure}[t]
\includegraphics[width=9cm]{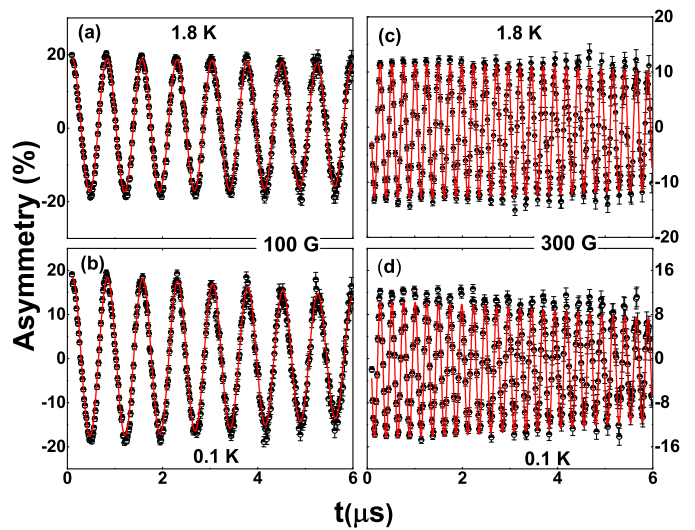}
\caption{\label{fig:TF_Asymmetry}(Color online) Transverse- field $\mu$SR spectra (one component) for LaPt$_2$Si$_2$ obtained at $T$ = 0.1 K and at $T$ = 1.8 K  in an applied magnetic field $H$ = 100 G [see (a)- (b)] $H$ = 300 G [see (c)-(d)] for field-cooled (FC) state. Solid red lines represent the fits to the observed spectra with Eq 1.}
\end{figure}

Fig. 1(a)-(b) and (c)-(d) show the TF-$\mu$SR precession signals obtained in FC condition under an applied field of 100~G and 300~G, respectively. It is quite evident that below $T_c$ the $\mu$SR precession signal decays with time in both the cases caused by the inhomogeneous field distribution of the flux-lattice emphasizing the fact that the sample is indeed in the superconducting mixed state. Observed TF-—$\mu$SR asymmetry spectra can be best fitted with an oscillatory decaying Gaussian function which is given by

\begin{equation}
\begin{gathered}
G_{TF}~(t)= A_0 \cos(2\pi\nu_1t +\phi)\exp\left(-\frac{\sigma_{tot}^2t^2} {2}\right)\\
+ A_{BG}\cos(2\pi\nu_2t +\phi)
\end{gathered}
\label{eq:1}
\end{equation}

\noindent where $\nu_1$ and $\nu_2$ represents the frequencies of muon precession signal originating from the superconducting fraction of the sample and the background due to  sample holder, respectively. $A_0$ and $A_{BG}$ are the muon initial asymmetries associated with the sample and background respectively, $\sigma_{tot}$ is the total sample relaxation rate and $\phi$ is the initial phase offset. Fitting of the observed spectra with Eq. 1 is presented by the solid red line in Figure 1(a)-(d). Considering the information related to the superconducting gap structure, the first term in Eq 1 is most important, as below $T_c$ it gives the total sample relaxation rate $\sigma_{tot}$ which contains contributions from the vortex lattice ($\sigma_{sc}$) and nuclear dipole moments ($\sigma_{nm}$), which is expected to be constant over the entire temperature range ($i.e.$ above and below $T_c$). $\sigma_{tot}$ is related to $\sigma_{sc}$ and $\sigma_{nm}$ by the relation $\sigma_{tot}$ = $\sqrt{\sigma_{sc}^2 + \sigma_{nm}^2}$. Thus, the contribution due to the vortex lattice, $\sigma_{sc}$, was obtained by quadratically subtracting the background nuclear dipolar relaxation rate obtained from the fitting of the spectra measured above $T_c$.

\begin{figure}[t]
\includegraphics[width=\linewidth]{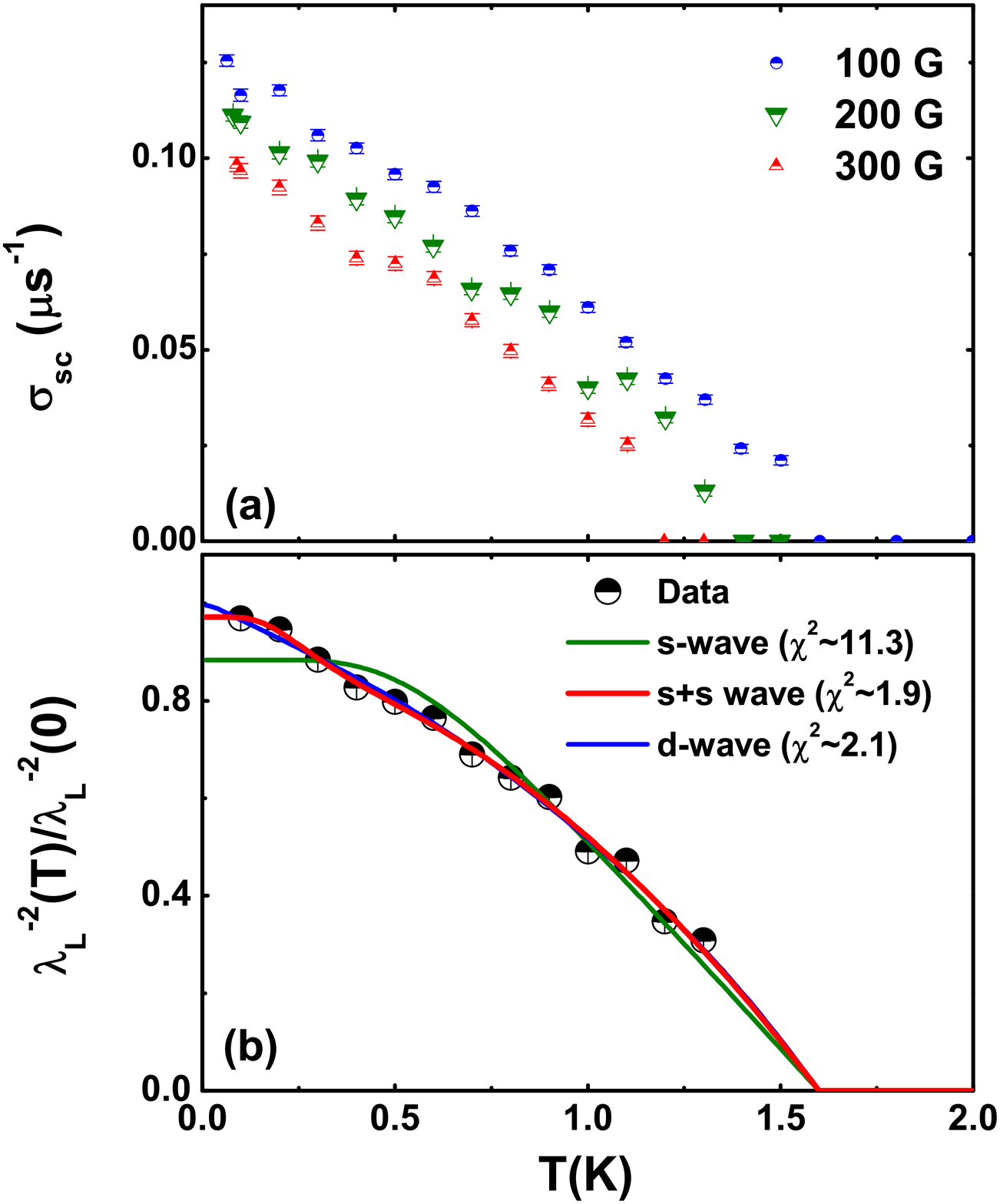}
\caption{\label{fig:Sigma_and_Gap}(Color online) (a) Temperature dependence of the muon depolarization rate $\sigma_{sc}$ of LaPt$_2$Si$_2$ measured under applied magnetic field of 100~G, 200~G and 300~G in field cooled (FC) condition. (b)Variation of $\lambda_L^{-2}(T)/\lambda_L^{-2}(0)$ as a function of temperature. The lines are fit to the data using an isotropic $s$-wave model, linear combination of two $s$ waves ($i.e$ $s+s$ wave) and $d$ wave with line nodes.}
\end{figure}

\noindent We obtain the magnetic field  and temperature dependence of  $\sigma_{sc}(T,H)$ by fitting Eq. (1) to the $\mu$SR time dependence asymmetry spectra. Figure 2(a) depicts the temperature dependence of $\sigma_{sc}$ for three different applied fields. After that we used the numerical Ginzburg-Landau model developed by Brandt,~\cite{Brandt}

\begin{equation}
\begin{gathered}
\sigma_{sc} [\mu s^{-1}] = 4.83 \times 10^4(1-H/H_{c2})\\ \times [1+1.21\sqrt{(1-H/H_{c2})^3}]\lambda_L^{-2} [nm]
\end{gathered}
\label{eq:2}
\end{equation}

to fit the field-dependent depolarization rate $\sigma_{sc} (H)$, and estimate two important superconducting order parameters, i.e., the London penetration depth $\lambda_L$ and the upper critical field $H_{c2}$. This model presume that $\lambda_L$ is field independent.  Now $\sigma_{sc}$ is directly related to the magnetic penetration depth ($\lambda_L$) which is associated with the superconducting gap structure. Therefore, $\sigma_{sc}$ can be modeled with the superconducting gap by the relation\cite{Prozorov}

\begin{equation}
\frac{\sigma_{sc}(T)} {\sigma_{sc}(0)} = \frac{\lambda_L^{-2}(T)} {\lambda_L^{-2}(0)} = 1 + 2 \left\langle \int_{\Delta}^{\infty}\int_{0}^{2\pi}\frac{\delta f}{\delta E} \frac{E~dEd\varphi}{\sqrt{E^2-\Delta_k^2}}\right\rangle_{FS}
\label{eq:2}
\end{equation}

\noindent where $f$ is the Fermi function given by $f$ = [1~+~$\exp (E/k_BT)$]$^{-1}$, the brackets $\langle \rangle_{FS}$ signifies the averaging over the Fermi surface and $\Delta$ represents the superconducting gap. This gap $\Delta$ which is a function of temperature and the azimuthal angle ($\varphi$) along the Fermi surface can be described as $\Delta(T, \varphi) = \Delta_0 \delta(T/T_c)~g(\varphi)$. Here, the temperature dependence of the of the superconducting gap is approximated by the relation~\cite{Carrington}~$\delta(T/T_c) = \tanh\left\{1.82[1.018*(T_c/T-1)]^{0.51}\right\}$. The spacial dependence $g(\varphi)$ = 1 for $s$ wave and $ \mid\cos(2\varphi)\mid$ for $d$ wave model with line nodes\cite{Prozorov, Carrington, Gross}. We have analyzed $\lambda_L^{-2}(T)/\lambda^{-2}_L(0)$ data estimated from the TF-$\mu$SR data analysis of 100, 200 and 300~G as shown in Fig.2 (b) using Eq.3. We have considered three models in our analysis: an isotropics ($s$ wave) gap model, a combination of two $s$ waves with different gaps ($i.e.$ $s+s$ wave model)  and $d$ wave model with line nodes.      

We have summarized superconducting gap parameter values obtained after fitting with different models in Table I. It can be seen from Fig 2 (b) that both $d$ wave model and $s+s$ wave model replicates the observed data quite well. But this apparent dilemma can be resolved by having a close look on the goodness of fitting which suggests that $s+s$ model give the lowest value of $\chi^2$ (for $s$ wave, $d$ wave and $s+s$ wave models $\chi^2$ = 11.3, 2.1 and 1.9 respectively) indicating the best fit to the observed data. It should be emphasized that our analysis\cite{Das} of electronic part of the specific heat (see Supplemental Material) is also suggestive of a $s+s$ wave gap structure in LaPt$_2$Si$_2$. Hence, analysis of specific heat and $\mu$SR data conjointly hint towards multigap superconductivity in LaPt$_2$Si$_2$. Moreover, NMR investigation on isostructural SrPt$_2$Si$_2$ shows a Hebel-Slichter coherence peak of 1/$T_1$ below $T_c$ \cite{Kawasaki} which indicates isotropic gap structure in the system. Hence, observation of two gaps in LaPt$_2$Si$_2$ is quite remarkable among the existing members of CaBe$_2$Ge$_2$- type structure exhibiting coexistence of CDW and superconductivity. However, recent reports discussing NMR studies \cite{Kubo, Aoyama} on LaPt$_2$Si$_2$ are limited down to 5~K. So, NMR investigations probing the superconducting state in LaPt$_2$Si$_2$ will be worthwhile.

\par

Now, considering London theory\cite{Ashcroft} $\lambda_L(0)$ can be related to the effective quasi-particle mass ($m^*$) and the superfluid density ($n_s$) by the relation $\lambda_L (0) = m^*c^2/ 4\pi n_s e^2$ where $m^*$ = (1+ $\lambda_{el-ph}$)$m_e$. The value of electron phonon coupling constant $\lambda_{el-ph}$ which can be derived from McMilan's relation, was already estimated to be 0.53 in our earlier report \cite{Ritu2}. Taking this value of $\lambda_{el-ph}$, we estimated $n_s$  for different models which have been presented in Table I. 

\begin{table}
	\caption{\label{tab:Hc fit}: Superconducting parameters obtained by fitting $\mu$SR data with different models.}
	\begin{ruledtabular}
		\begin{tabular}{c c c c c c c}
			Model & Gap & $\lambda_L(0)$  &$n_s$$ \times10^{27}$ &$\chi^2$  \\
			&($\Delta_0/k_BT_c$)& (nm)& ($m^{-3}$)&\\
			\hline\\
			$s-$wave   & 1.523 & 292.86 & 4.99 & 11.3\\
			$d-$wave   & 2.278   & 275.95 & 5.63 &  2.1 \\
			$s+s-$wave & 1.846 & 279.36 & 5.49 & 1.9\\
			      &  0.475&&\\

		\end{tabular}
	\end{ruledtabular}
\end{table}

\begin{figure}[t]
	\includegraphics[width=\linewidth]{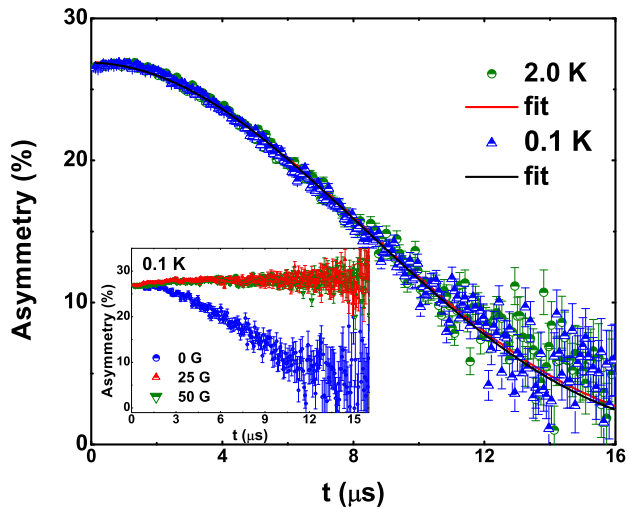}
	\caption{\label{fig:Lambda_and_ZF}(Color online) ZF $\mu$SR spectra of LaPt$_2$Si$_2$ recorded at $T$ = 0.1~K and 2~K. Solid line represents fitting of the observed spectra with KT function (see text). Inset: Comparison of LF spectra measured for $H$ = 0, 25~G and 50~G at 0.2~K. }
\end{figure}

ZF-$\mu$SR spectra for temperatures above and below $T_c$ has been presented in Fig.3. Here, the muon-spin relaxation, observed in the ZF-$\mu$SR spectra, is possibly due to static, randomly oriented local fields associated with the nuclear moments at the muon site. Observed ZF-$\mu$SR spectra can be well illustrated using a damped Gaussian Kubo-Toyabe (KT) function, $G_{ZF}(T)~= A_1 G_{KT}\exp(-\Lambda t) + A_{BG}$, where $A_1$ is the initial asymmetry, $A_{BG}$ is the temperature independent background originating due to muons stopping in the sample holder, $\Lambda$ is the electronic relaxation rate and $G_{KT}$ is the Gaussian Kubo-Toyabe (KT) function which is expected from an isotropic Gaussian distribution of randomly oriented static (or quasistatic) local fields at the muon sites and is defined as\cite{KT}, $G_{KT} = \left[\frac{1}{3}+ \frac{2}{3} (1-\sigma_{KT}^2t^2) \exp\left( -\frac{\sigma_{KT}^2t^2}{2}\right)\right]$, with $\sigma_{KT}$ being the muon depolarization rate. It is evident from Fig. 3 that ZF-$\mu$SR spectra collected above and below $T_c$ show no noticeable change in the relaxation rates. This observation suggests that the time-reversal symmetry remains preserved upon entering the superconducting state. On the other hand, small application of a longitudinal magnetic field of just 25~G (see inset of Fig. 3) confiscates any relaxation due to nuclear static fields and is sufficient to fully decouple the muons from this relaxation channel. 

\subparagraph*{}

In summary, we have investigated the superconducting gap structure of a CDW SC LaPt$_2$Si$_2$ having $T_c$ = 1.6~K using TF, ZF and LF muon spin rotation/ relaxation measurements. We have determined the temperature dependence of muon depolarization rate due to the superconducting sample, $\sigma_{sc}$, by analyzing TF $\mu$SR data. Our analysis suggest that the superconducting gap structure in LaPt$_2$Si$_2$ can be best fitted with two gap $s$ wave model ($s+s$ wave) rather than an isotropic $s$-wave model. This conclusion is in agreement with the specific heat analysis which also indicates multigap superconductivity in LaPt$_2$Si$_2$. On the other hand, ZF data do not show any indication of TRS breaking. Further investigations using other microscopic techniques such as tunnel diode oscillator or scanning tunneling microscopy on good quality single crystals probing the presence of multiple superconducting gaps will be quite interesting.

\subparagraph*{}

We thank Dr. R. Kulkarni, Prof. S.K. Dhar, Prof. A. Thamizhavel and Mr. K. Panda for cooperation and useful discussions. We also thank Dr. Biswanath Samanataray for help with the analysis of specific heat data. DD would like to thank Newton funding for providing travel support to carry out $\mu$SR experiments. AB would like to acknowledge DST India, for Inspire Faculty Research Grant (DST/INSPIRE/04/2015/000169). DTA  would like to thank CMPC-STFC, grant number CMPC-09108, for financial support.


\begin{thebibliography}{44}

\bibitem{Cooper}
L.N. Cooper, Phys. Rev. {\bf 104}, 1189 (1956).

\bibitem{Bardeen1}
J. Bardeen, L.N. Cooper and J.R. Schrieffer, Phys. Rev, {\bf 106}, 162 (1957).

\bibitem{Bardeen2}
J. Bardeen, L.N. Cooper and J.R. Schrieffer, Phys. Rev. {\bf 108}, 1175 (1957).

\bibitem{Bednorz}
J. G. Bednorz and K. A. Muller, Z. Phys. B {\bf 64}, 189 (1986).

\bibitem{Steglich}
F. Steglich, J. Aarts, C. D. Bredl, W. Lieke, D. Meschede,W. Franz, and H. Sch¨afer, Phys. Rev. Lett. {\bf 43}, 1892 (1979).

\bibitem{White}
B.D. White, J.D. Thompson, M.B. Maple, Physica C {\bf 514}, 246 (2015).

\bibitem{Kamihara}
Y. Kamihara, T. Watanabe, M. Hirano and H. Hosono, J. Am. Chem. Soc. {\bf 130}, 3296 (2008).

\bibitem{Johnston}
D. C. Johnston, Adv. Phys. {\bf 59}, 803(2010).

\bibitem{Takahashi}
H. Takahashi, K. Igawa, K. Arii, Y. Kamihara, M. Hirano and H. Hosono, Nature (London) {\bf 453}, 376 (2008).

\bibitem{Jeevan}
H. S. Jeevan, Z. Hossain, D. Kasinathan, H. Rosner, C. Geibel and P. Gegenwart, Phys. Rev. B {\bf 78}, 92406(2008).

\bibitem{Imre}
A. Imre, A. Hellmann, G. Wenski, J. Graf, D. Johrendt and A. Mewis, Z. Anorg. Allg. Chem. {\bf 633}, 2037 (2007).

\bibitem{Kudo}
K. Kudo, Y. Nishikubo, and M. Nohara, J. Phys. Soc. Jpn. {\bf 79}, 123710 (2010).

\bibitem{Nagano}
Y. Nagano, N. Araoka, A. Mitsuda, H. Yayama, H. Wada, M. Ichihara, M. Isobe, and Y. Ueda, J. Phys. Soc. Jpn. {\bf 82}, 64715 (2013)

\bibitem{Gruner}
T. Gruner, D. Jang, Z. Huesges, R. Cardoso-Gil, G. H. Fecher, M. M. Koza, O. Stockert, A. P. Mackenzie, M. Brando and Christoph Geibel, Nature Physics, {\bf 13}, 967 (2017).

\bibitem{Ritu1}
R. Gupta, U. B. Paramanik, S. Ramakrishnan, K. P. Rajeev and Z. Hossain, J.Phys.:Condens. Matter {\bf 28}, 195702 (2016).

\bibitem{Ritu2}
R. Gupta, S. K. Dhar, A. Thamizhavel, K. P. Rajeev and Zakir Hossain, J.Phys.:Condens. Matter {\bf 29}, 255601 (2017).

\bibitem{Anand}
V K Anand, Z Hossain and C Geibel, J.Phys.:Condens. Matter {\bf 19}, 486207 (2007)

\bibitem{Kumar}
M. Kumar, V. K. Anand, C. Geibel, M. Nicklas and Z. Hossain, Phys. Rev. B, {\bf 81}, 125107 (2010).


\bibitem{Eisenmann}
B. Eisenmann, N. May, Wiking M\"{u}ller and Herbert Sch\"{a}fer, Zeitschrift für Naturforschung B, {\bf 27}, 1155 (1972).

\bibitem{Kawasaki}
S. Kawasaki, Y. Tani, T. Mabuchi, K Kudo, Y. Nishikubo,D. Mitsuoka, M. Nohara, and G.-q. Zheng, Phys. Rev. B, {\bf 91}, 60510 (2015).

\bibitem{Kimura}
N. Kimura, K. Ito, K. Saitoh, Y. Umeda, H. Aoki, and T. Terashima, Phys. Rev. Lett. {\bf 95}, 247004 (2005).

\bibitem{Pecharsky}
V. K. Pecharsky, L. L. Miller, K. A. Gschneidner, Phys. Rev. B {\bf 58}, 497 (1998).

\bibitem{Hillier}
A. D. Hillier, J. Quintanilla, and R. Cywinski, Phys. Rev. Lett. {\bf 102}, 117007 (2009).

\bibitem{Bauer}
E. Bauer, R. T. Khan, H. Michor, E. Royanian, A. Grytsiv, N. Melnychenko-Koblyuk, P. Rogl, D. Reith, R. Podloucky, E. W. Scheidt, W. Wolf, and M. Marsman, Phys. Rev. B {\bf 80}, 064504 (2009).

\bibitem{Smidman}
M. Smidman, A. D. Hillier, D. T. Adroja, M. R. Lees, V. K. Anand, R. P. Singh, R. I. Smith, D. M. Paul, and G. Balakrishnan, Phys. Rev. B {\bf 89}, 094509 (2014).

\bibitem{Luke}
G. M. Luke, Y. Fudamoto, K. M. Kojima, M. I. Larkin, J. Merrin, B. Nachumi, Y. J. Uemura, Y. Maeno, Z. Q. Mao, Y. Moriet al., Nature (London) {\bf 394}, 558(1998).

\bibitem{gml} G. M. Luke, A. Keren, L. P. Le, W. D. Wu, Y. J. Uemura, D. A. Bonn, L. Taillefer, and J. D. Garrett, Phys. Rev. Lett. {\bf 71}, 1466 (1993).

\bibitem{rhh} R. H. Heffner, J. L. Smith, J. O.Willis, P. Birrer, C. Baines, F. N. Gygax, B. Hitti, E. Lippelt, H. R. Ott, A. Schenck et al., Phys. Rev. Lett. {\bf 65}, 2816 (1990).

\bibitem{ya} Y. Aoki, A. Tsuchiya, T. Kanayama, S. R. Saha, H. Sugawara, H. Sato, W. Higemoto, A. Koda, K. Ohishi, K. Nishiyama {\it et al.,} Phys. Rev. Lett. {\bf 91}, 067003 (2003).

\bibitem{am} A. Maisuradze, W. Schnelle, R. Khasanov, R. Gumeniuk, M. Nicklas, H. Rosner, A. Leithe-Jasper, Y. Grin, A. Amato, and P. Thalmeier, Phys. Rev. B {\bf 82}, 024524 (2010).

\bibitem{ad1}  A. D. Hillier, J. Quintanilla, and R. Cywinski, Phys. Rev. Lett. {\bf 102}, 117007 (2009).

\bibitem{ad2}  A. D. Hillier, J. Quintanilla, B. Mazidian, J. F. Annett, and R. Cywinski, Phys. Rev. Lett. {\bf 109}, 097001 (2012).

\bibitem{rps}  R. P. Singh, A. D. Hillier, B. Mazidian, J. Quintanilla, J. F. Annett, D. M. Paul, G. Balakrishnan, and M. R. Lees, Phys. Rev. Lett. {\bf 112}, 107002 (2014).

\bibitem{Adroja1}
D. T. Adroja, A. Bhattacharyya, M. Telling, Yu. Feng, M. Smidman, B. Pan, J. Zhao, A. D. Hillier, F. L. Pratt and A. M. Strydom, Phys. Rev. B {\bf 92}, 134505 (2015).

\bibitem{Adroja2}
D. T. Adroja, , A. Bhattacharyya, M. Smidman, A. Hillier,Y. Feng, B. Pan, J. Zhao, M. R. Lees, A. Strydom and P. K. Biswas, J. Phys. Soc. Jpn. {\bf 86} 44710 (2017).

\bibitem{Bhattacharyya1}
A. Bhattacharyya,D. T. Adroja, J. Quintanilla, A. D. Hillier, N. Kase, A. M. Strydom and J. Akimitsu, Phys. Rev. B, {\bf 91}, 060503(R) (2015).


\bibitem{Bhattacharyya2}
A. Bhattacharyya, D. Adroja, N. Kase, A. Hillier, J. Akimitsu and A. Strydom, Sci. Rep., {\bf 5}, 12926 (2015).

\bibitem{Quintanilla}
J. Quintanilla, A. D. Hillier, J. F. Annett, and R. Cywinski,Phys.Rev. B {\bf 82}, 174511 (2010).

\bibitem{Bauer2}
E.Bauer, C.Sekine,U.Sai,P.Rogl,P.K.Biswas,and A.Amato, Phys. Rev. B {\bf 90}, 054522 (2014).

\bibitem{Kubo}
T. Kubo, Y. Kizaki, H. Kotegawa, H. Tou, Y. Nagano,N. Araoka, A. Mitsuda and H. Wada, JPS Conf. Proc. {\bf 3}, 17031 (2014).


\bibitem{Kim}
S. Kim, K. Kim and  B. I. Min, Sci. Rep., {\bf 5}, 15052 (2015).

\bibitem{Hase}
I. Hase and T. Yanagisawa, Physica C, {\bf 484}, 59 (2013).

\bibitem{Sonier}
J. E. Sonier, J. H. Brewer, and R. F. Kiefl,Rev. Mod. Phys.{\bf 72}, 769 (2000).

\bibitem{Das}
See Supplemental Material for detail of sample characterization and analysis of heat capacity data.

\bibitem{Lee}
S. L. Lee, S. H. Kilcoyne, and R. Cywinski, Muon Science: Muons in Physics, Chemistry and Materials; SUSSP Publications and IOP Publishing, Bristol, U.K. (1999).


\bibitem{Pratt}
F. L. Pratt, Physica B, {\bf 289}, 710 (2000)

\bibitem{Brandt} 
E. H. Brandt, Phys. Rev. B {\bf 68}, 054506 (2003).

\bibitem{Prozorov}
R. Prozorov and R. W. Giannetta,Supercond. Sci. Technol. {\bf 19}, R41 (2006).

\bibitem{Carrington}
A. Carrington and F. Manzano,Physica C {\bf 385}, 205 (2003)

\bibitem{Gross}
F. Gross, B. S. Chandrasekhar, D. Einzel, K. Andres, P. J. Hirschfeld, and H. R. Ott,Z. Phys. B: Condens. Matter {\bf 64}, 175(1986).


\bibitem{Aoyama}
T. Aoyama, T. Kubo, H. Matsuno, H. Kotegawa, H. Tou, A. Mitsuda, Y. Nagano, N. Araoka, H. Wada and Y. Yamada, Journal of Physics: Conf. Series {\bf 807}, 062002 (2017) .

\bibitem{Ashcroft}
N. W. Ashcroft and N. D. Mermin, Solid State Physics, p. 738; Saunders College, Philadelphia, (1976).

\bibitem{KT}
R. Kubo and T. Toyabe, In Magnetic Resonance and Relaxation, ed. R. Blinc (North-Holland, Amsterdam, 1967) p. 810; T. Toyabe, M.S. thesis, University of Tokyo (1966).


\end{thebibliography}
\end{document}


\title{Multigap Superconductivity in a charge density wave superconductor LaPt$_2$Si$_2$}
\author{Debarchan Das}
\thanks{Present address: Institute of Low Temperature and Structure Research, Polish Academy of Sciences, Wroc{\l}aw 50-950, Poland}
\affiliation{Department of Physics, Indian Institute of Technology, Kanpur 208016, India}
\author{Ritu Gupta}
\affiliation{Department of Physics, Indian Institute of Technology, Kanpur 208016, India}
\author{A. Bhattacharyya}
\email{amitava.bhattacharyya@rkmvu.ac.in}
\address{Department of Physics, Ramakrishna Mission Vivekananda Educational and Research Institute, Belur Math, Howrah 711202, West Bengal, India}
\author{P. K. Biswas}
\affiliation{ISIS Facility, STFC, Rutherford Appleton Laboratory,Chilton, Oxfordshire, OX11 0QX, United Kingdom}
\author{D. T. Adroja}
\affiliation{ISIS Facility, STFC, Rutherford Appleton Laboratory,Chilton, Oxfordshire, OX11 0QX, United Kingdom}
\affiliation{Highly Correlated Matter Research Group, Physics Department, University of Johannesburg, PO Box 524, Auckland Park 2006, South Africa}
\author{Z. Hossain}
\email{zakir@iitk.ac.in}
\affiliation{Department of Physics, Indian Institute of Technology, Kanpur 208016, India}

\maketitle

In this supplementary note we present the detail of sample characterization performed on LaPt$_2$Si$_2$ sample. Phase purity of the polycrystalline sample was checked by powder x-ray diffraction (XRD) using Cu-K$_\alpha$ radiation. XRD pattern obtained at room temperature was analyzed by Rietveld refinement using FullProf software\cite{Carvajal}. Fig 1. represents powder XRD pattern  along with Rietveld refinement profile which confirms the single phase nature of the sample. Lattice parameters [$a$ = 4.291(1)~\AA~and $c$ = 9.836(2)~\AA] obtained from refinement are in good agreement with our earlier report \cite{Ritu1}. Sample homogeneity was checked by performing different metallographic experiments like scanning electron microscopy (SEM) and energy dispersive x-ray (EDX) analysis. Furthermore, to characterize the physical properties of this sample, we measured its electrical resistivity in a Quantum Design PPMS system equipped with $^3$He option. The temperature dependence of electrical resistivity, $\rho(T)$, measured in the temperature range of 0.4~K to 300~K has been depicted in Fig 2(a) which confirms the coexistence of superconductivity and charge density wave (CDW) in this particular sample. The transition temperatures ($T_c$ = 1.6~K and $T_{CDW}$ = 112~K) are similar to our earlier report \cite{Ritu1}.

\begin{figure}[htb!]
	\includegraphics[width=8.7cm, keepaspectratio]{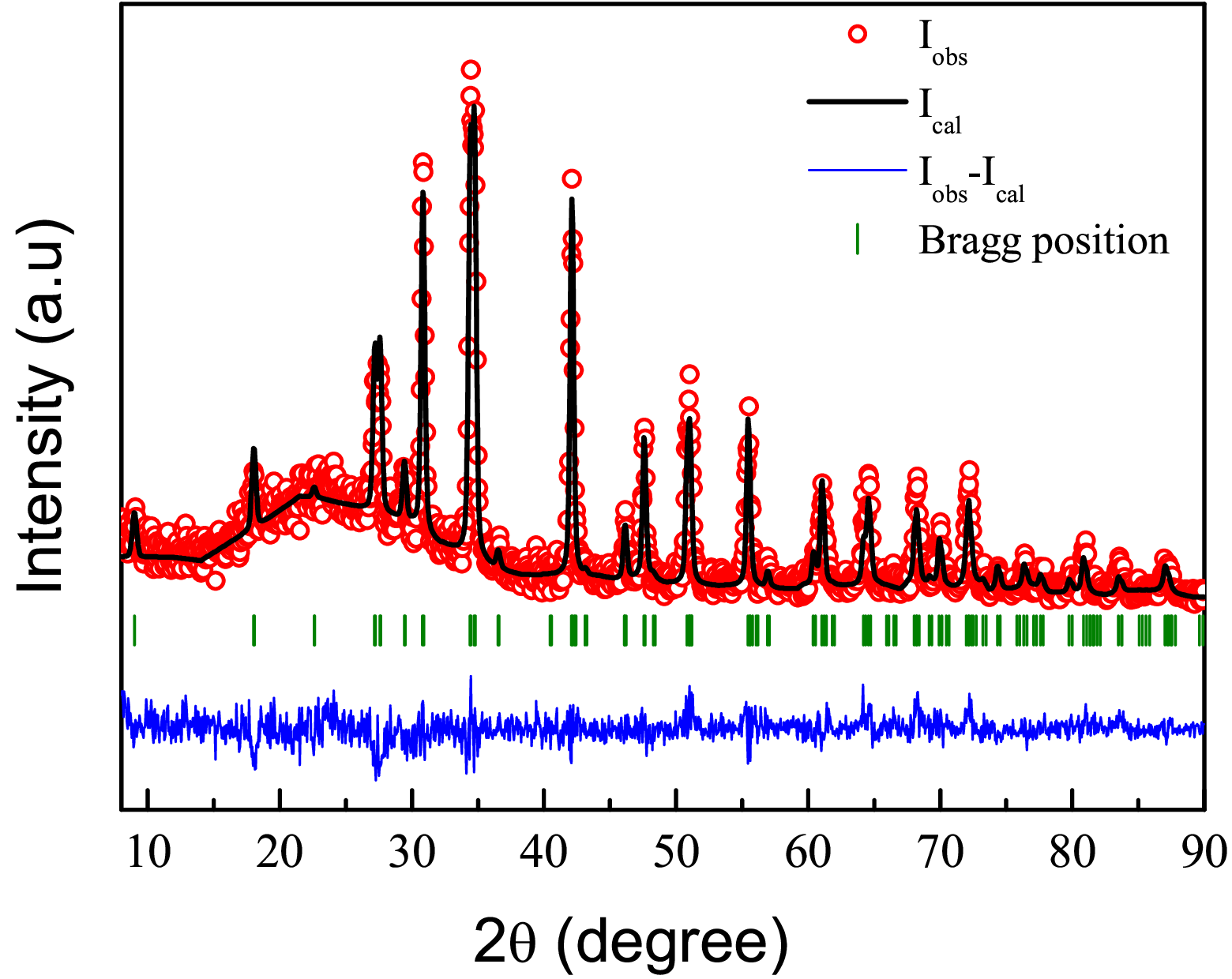}
	\caption{\label{fig:RT_and_HC}(Color online) The powder x-ray diffraction pattern of LaPt$_2$Si$_2$ recorded at room temperature. The solid line through the experimental points is the Rietveld refinement profile. The short vertical bars represent the Bragg peak positions. The lowermost curve (blue line) represents the difference between the experimental and calculated intensities. Goodness of fitting, $\chi^2$ = 1.29}
\end{figure}

Moreover, in order to have an idea about the superconducting gap structure of LaPt$_2$Si$_2$, we analyzed the specific heat data obtained on single crystalline sample [shown in Fig. 2(b)] as reported previously by us \cite {Ritu2}. The electronic part of the specific heat $C_{el}$ is related to the entropy $S_{sc}$ in  superconducting state by the relation\cite{Nakajima, Chen}
\begin{equation}
C_{el}  = T \left(\frac{\delta S_{sc}}{\delta T}\right)
\label{eq:2}
\end{equation}
\noindent Within the formulation of BCS theory, $S_{sc}$ can be written as
\begin{equation}
S_{sc}  = - \frac{3\gamma_n}{\kappa_B\pi^3}\int_0^{2\pi}\int_0^{\infty}\left[(1-f) ln(1-f)+f lnf\right] d\varepsilon d\phi
\label{eq:3}
\end{equation}

\begin{figure}[htb!]
	\includegraphics[width=8.7cm, keepaspectratio]{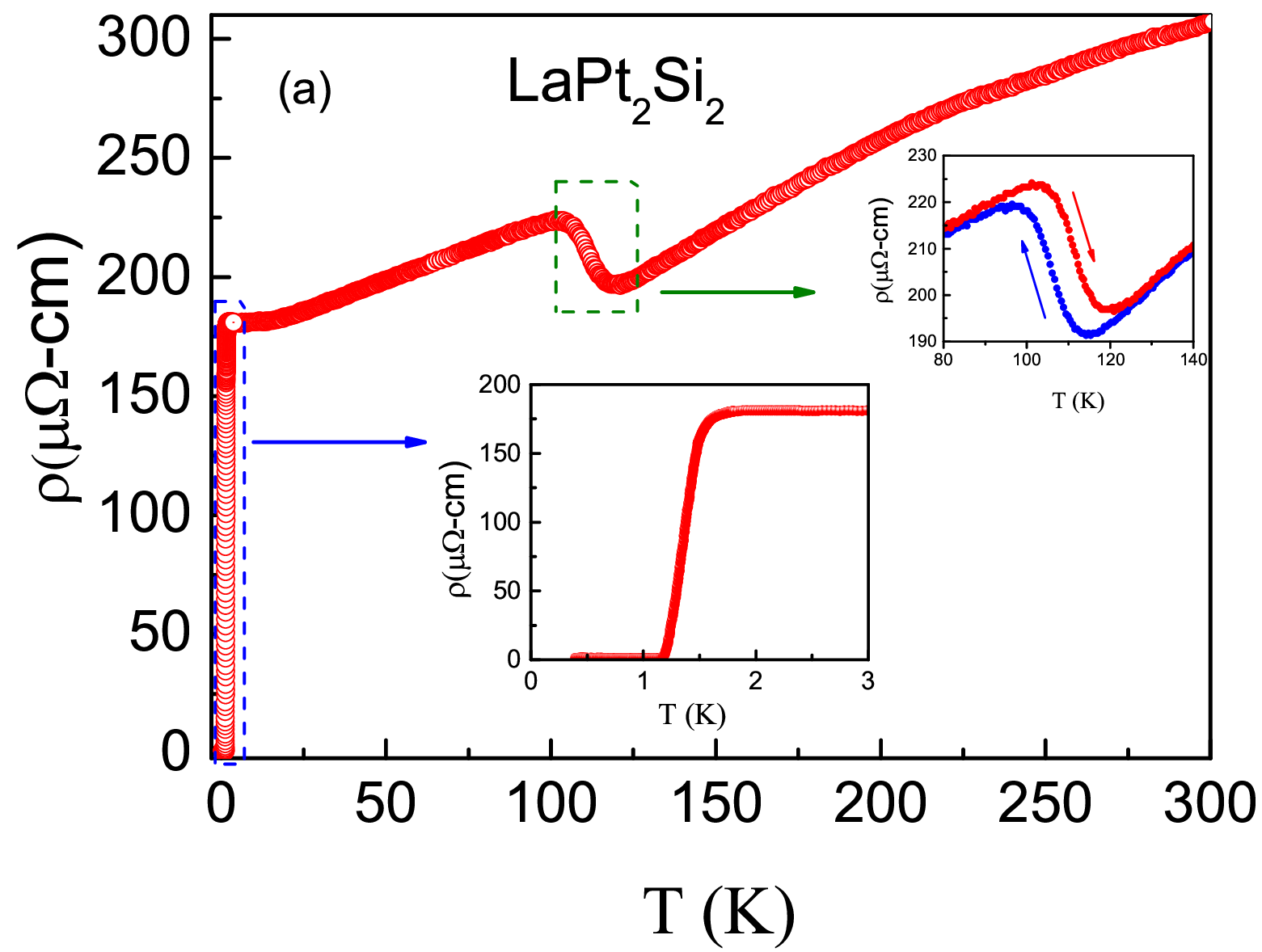}
	\includegraphics[width=8.3cm, keepaspectratio]{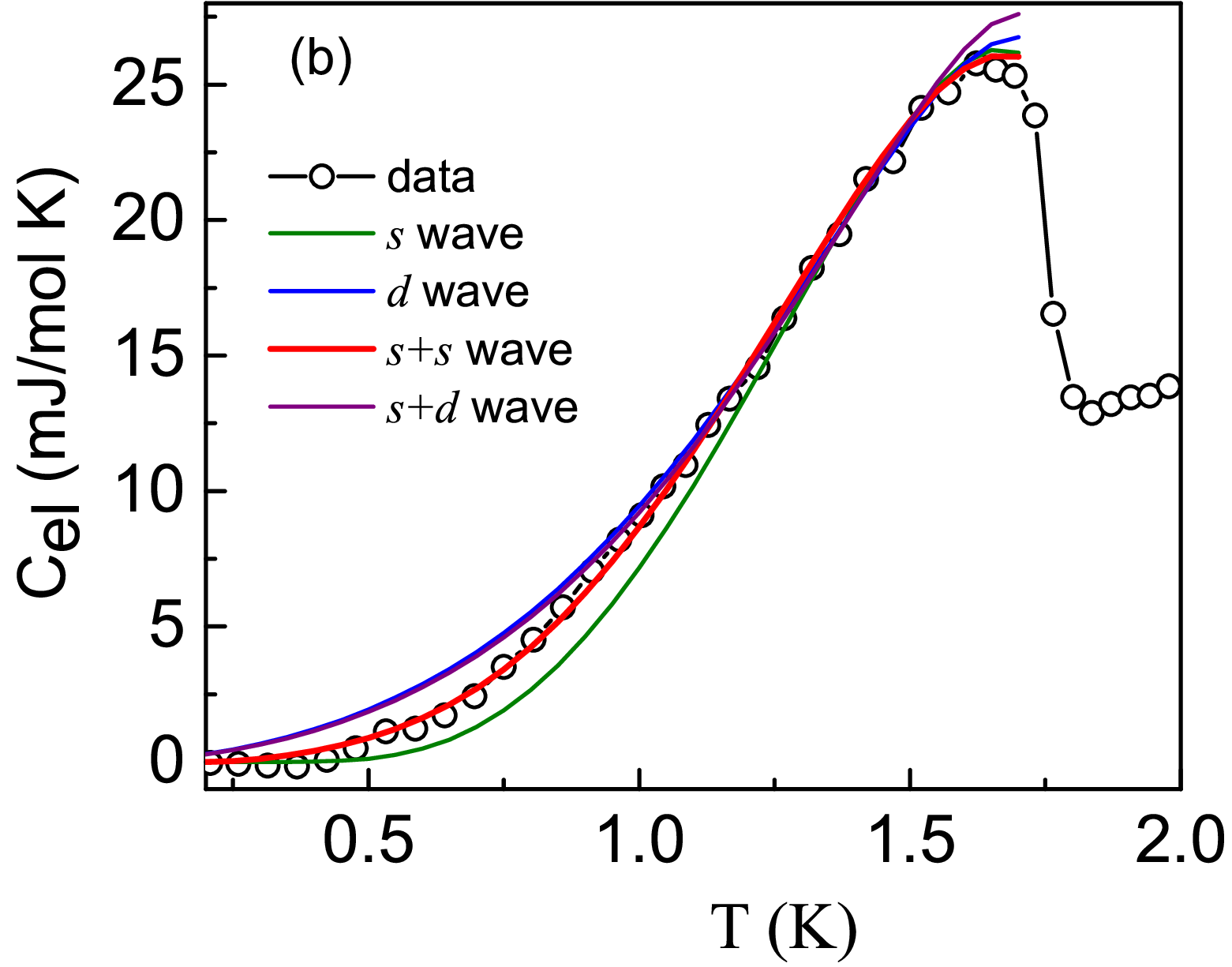}
	\caption{\label{fig:RT_and_HC}(Color online) (a) Temperature dependence of electrical resistivity. Inset images emphasize on the CDW transition and SC transition as indicated (b) Variation of the electronic part of specific heat measured on single crystalline sample. Solid lines correspond to the fitting of experimentally obtained data with different models as described in the text.}
\end{figure}

\begin{table}
\caption{\label{tab:Hc fit}: Parameters obtained after fitting the specific heat data with different models as discussed in the text}
\begin{ruledtabular}
\begin{tabular}{c c c c c c c}
Model & $\gamma$ & $\Delta/k_B T_c$ &$x$ \\
 &(mJ mol$^{-1}$ K$^{-2}$)\\
 \hline
 \\
 $s$   & 7.8&2.7&-\\
 $d$   & 8.1&3.6&-\\
 $s+s$ & $\gamma_1$ =8.7 , $\gamma_2$= 5.7&2.7 and 1.1& 0.78\\
 $s+d$ & $\gamma_s$ = 5.7, $\gamma_d$= 9.0&4.3 and 3.4& 0.80\\

\end{tabular}
\end{ruledtabular}
\end{table}
\noindent where $\gamma_n$ is the normal state Sommerfeld coefficient which is set as one of the free parameters in the fitting, $f$ represents quasiparticle occupation number given by $f$ = [1~+~$\exp (E/k_BT)$]$^{-1}$ having $E$ = $\sqrt{\varepsilon^2+\Delta^2(\phi,T)}$. The temperature dependence of gap function was defined using well established $\alpha$ model where gap function is usually scaled by a parameter $\alpha$, $\Delta$($\phi$, $T$) = $\alpha$~$\Delta$$_{BCS}$($\phi$, $T$). More details about the $\alpha$ can be found in the literature\cite{David, Muhlschlegel}. $\Delta$$_{BCS}$$ (\phi,$T$)$ is the BCS supercoducting gap function given by \cite{Prozorov,Carrington}
$\Delta_{BCS}(\phi,T) = g(\phi)~\tanh\left\{1.82[1.018*(T_c/T-1)]^{0.51}\right\}~\Delta$. Where $\Delta$ is the gap function at zero Kelvin. Combining Eq 1 and 2, we analyzed $C_{el}$ with different gap structure namely $s$ wave, $d$ wave, $s+s$ wave and $s+d$ wave. In case of simple $s$ wave, the angular dependence of gap function is given by $g(\phi)$ = 1 whereas $g(\phi)$ = $\cos (2\phi)$ for $d$ wave gap structure. In case of $s+s$ wave scenario we considered linear combination of two $s$ waves $i.e.$ $C_{el}$ = $x C_{s1}+(1-x)C_{s2}$ where $C_{s1}$ and $C_{s2}$ are expressed in the form of single $s$ wave with different gap values ($\Delta$). On the other hand, for $s+d$ wave, $C_{el}$ can be expressed as $C_{el}$ = $x C_{s}+(1-x)C_{d}$ where $C_{s}$ and $C_{d}$ are written in the form of $s$ and $d$ waves respectively with different $\Delta$. It is quite evident from Fig 2 (b) that $s+s$ wave model gives the best fit with the experimentally observed data supporting the evidence of superconductivity having two distinct gaps similar to MgB$_2$\cite{Bouquet}, SrPt$_2$As$_2$\cite{Xu}, TiNi$_2$Se$_2$\cite{Wang}, Sc$_5$Ir$_4$Si$_10$\cite{Tamegai} . In Table I, we have summarized different fitting parameters obtained for different models. Thus, our specific heat analysis together with $\mu$SR results, strongly support multigap nature of superconducting gap structure in LaPt$_2$Si$_2$.